\documentclass[conference]{IEEEtran}

\usepackage{graphicx}
\usepackage{amsthm,bm}
\usepackage{algorithmic}
\usepackage{algorithm}
\usepackage{array}
\usepackage{stfloats}
\usepackage{url}
\usepackage{amsfonts}

\begin{document}
%
\title{Geographical Scheduling for Multicast\\ Precoding in Multi-Beam Satellite Systems}

\author{\IEEEauthorblockN{Alessandro Guidotti}
\IEEEauthorblockA{Department of Electrical, Electronic,\\ and Information Engineering\\ University of Bologna\\ Bologna, 40134, Italy\\
Email: a.guidotti@unibo.it}
\and
\IEEEauthorblockN{Alessandro Vanelli-Coralli}
\IEEEauthorblockA{Department of Electrical, Electronic,\\ and Information Engineering\\ University of Bologna\\ Bologna, 40134, Italy\\
Email: a.guidotti@unibo.it}
}

\maketitle

\begin{abstract}
Current State-of-the-Art High Throughput Satellite systems provide wide-area connectivity through multi-beam architectures. Due to the tremendous system throughput requirements that next generation Satellite Communications (SatCom) expect to achieve, traditional 4-colour frequency reuse schemes are not sufficient anymore and more aggressive solutions as full frequency reuse are being considered for multi-beam SatCom. These approaches require advanced interference management techniques to cope with the significantly increased inter-beam interference both at the transmitter, \emph{e.g.}, precoding, and at the receiver, \emph{e.g.}, Multi User Detection (MUD). With respect to the former, several peculiar challenges arise when designed for SatCom systems. In particular, multiple users are multiplexed in the same transmission radio frame, thus imposing to consider multiple channel matrices when computing the precoding coefficients. In previous works, the main focus has been on the users' clustering and precoding design. However, even though achieving significant throughput gains, no analysis has been performed on the impact of the system scheduling algorithm on multicast precoding, which is typically assumed random. In this paper, we focus on this aspect by showing that, although the overall system performance is improved, a random scheduler does not properly tackle specific scenarios in which the precoding algorithm can poorly perform. Based on these considerations, we design a Geographical Scheduling Algorithm (GSA) aimed at improving the precoding performance in these critical scenarios and, consequently, the performance at system level as well. Through extensive numerical simulations, we show that the proposed GSA provides a significant performance improvement with respect to the legacy random scheduling.

\end{abstract}


%
\IEEEpeerreviewmaketitle

\section{Introduction}
During the last years, Satellite Communication (SatCom) systems evolved from the traditional single-beam architecture to a more advanced multi-beam deployment, aiming at improving the system throughput. State-of-theArt (SoA) High Throughput Satellite (HTS) systems in Ka-band provide connectivity to single regions, \emph{e.g.}, Europe, by means of a very large number of beams, \emph{e.g.}, more than $100$, \cite{Intro6}. The basic principle behind this system architecture is the frequency reuse concept, which is well-known in the terrestrial community, with SoA HTS systems typically exploiting 4-colour frequency reuse schemes. However, when aiming at providing terabit connectivity through satellite systems, different solutions are needed to increase the system throughput, \emph{i.e.}, increasing either the available spectrum or the spectral efficiency at system level. On the one hand, several activities focused on the former by means of on advanced spectrum sharing technologies, \cite{Intro1}-\cite{Intro7}. On the other hand, more aggressive frequency reuse schemes, as \emph{full frequency reuse}, can be targeted in order to fully exploit the available licensed bandwidth and target much larger throughput levels. In this context, the overall system performance is strongly limited by the interference between adjacent beams. It is, thus, of paramount importance to implement advanced interference mitigation techniques at the receiver, \emph{e.g}, Multi-User Detection (MUD), \cite{Intro8}, or at the transmitter, \emph{i.e.}, precoding, \cite{Prec1}-\cite{ESAMIMO}.

The success of multi-user Multiple-Input Multiple-Output (MU-MIMO) techniques in terrestrial communications, together with the introduction of the super-frame structure in the DVB-S2X standard, \cite{DVBS2X}, led the Satellite Communications community to assess and implement precoding techniques in multi-beam HTS systems. In \cite{Prec1}-\cite{Prec3}, the authors proposed both Zero-Forcing (ZF) and Minimum Mean Square Error (MMSE) algorithms and some considerations on the main implementation challenges as the impact of partial Channel State Information (CSI) at the transmitter (CSIT). In \cite{Prec11,Prec6}, practical challenges related to the implementation of precoding to DVB-S2X HTS systems are discussed, \emph{e.g.}, framing and multiple gateways. In \cite{Prec5,Prec7}, the authors provide a review of several precoding techniques and propose an optimisation of the linear precoding design, with linear and non-linear power constraints, \cite{Prec5}. The authors of \cite{Prec4} implemented the Tomlinson-Harashima precoding (THP) by also taking into account the beam gain. In \cite{Prec10}, the authors assessed the performance of linear beamforming in terms of satisfying specific traffic demands. 

Building on these works, the SatCom community has been intensifying the work on precoding-based satellite systems. In particular, since SatCom systems provide large single-link data rates to the users, multicast precoding techniques have gained momentum. In this case, in order to enhance the overall system level efficiency, multiple users are multiplexed into the same PHY codeword. Within this framework, traditional single-user (unicast) precoding is not always applicable and multicast precoding techniques can provide an efficient solution as shown in several recent works, \cite{Taricco}-\cite{UniBo_Aerospace}. In \cite{Taricco}, the authors proposed the design of multicast precoding for satellite systems based on a regularised channel inversion and a geographical-based grouping of users in the same FEC codeword. The computation of the precoding matrix is based on the pragmatic approach proposed in \cite{CTTCprec}, in which the authors jointly design the linear precoding and ground-based beamforming at the gateway. The authors of \cite{Prec8} discuss on the implementation of linear precoding techniques to multi-beam broadband fixed satellite communications by also also introducing some preliminary consideration on the issues related to users grouping. In \cite{Prec12}, the optimisation problem of multicast precoding with per-antenna power constraints is addressed. In \cite{Prec13}, the authors focus on framing multiple users per transmission and on the per-antenna transmit power limitations and propose a solution for frame-based precoding based on the principles of physical layer multicasting to multiple co-channel groups under per-antenna constraints. In \cite{Prec14}, a two stage linear precoding is proposed to lower the complexity in the ground segment, under the presence of non-ideal CSI as well. In \cite{UniBo_Aerospace}, we extensively addressed the grouping of users in multicast precoding as a clustering problem, providing an insight on the most critical design aspects to be considered with different power constraints on MMSE precoding, similarity metrics, and algorithm complexity.

The valuable contributions provided in the literature showed significant gains in terms of average spectral efficiency at system level. However, these analyses did not take into account aspects related to the system scheduler, which have a strong impact on the precoding performance. In particular, as highlighted above, the research efforts have been mainly focused on the implementation of more advanced precoding algorithms and/or different grouping strategies for multicast precoding. While these aspects are relevant for the precoding implementation, they do not capture the whole picture since also the choice of the users to be served in the same time frame has a strong impact on the system performance. 

In this paper, we move from the work performed in \cite{UniBo_Aerospace} with respect to the users' clustering and provide an analysis of the system scheduler. In the literature, the users are always assumed to be randomly selected across the system beams. However, as we will show in the next sections, this choice can lead to specific scenarios in which the precoded system has a worse performance with respect to the non-precoded one. Moving from this analysis, we propose and design a Geographical Scheduling Algorithm (GSA), which aims at improving the performance of multicast and unicast precoding in these critical scenarios, which results in a significant overall gain at system level as well.

\begin{figure}[t]
\centering
\includegraphics[width=0.5\textwidth]{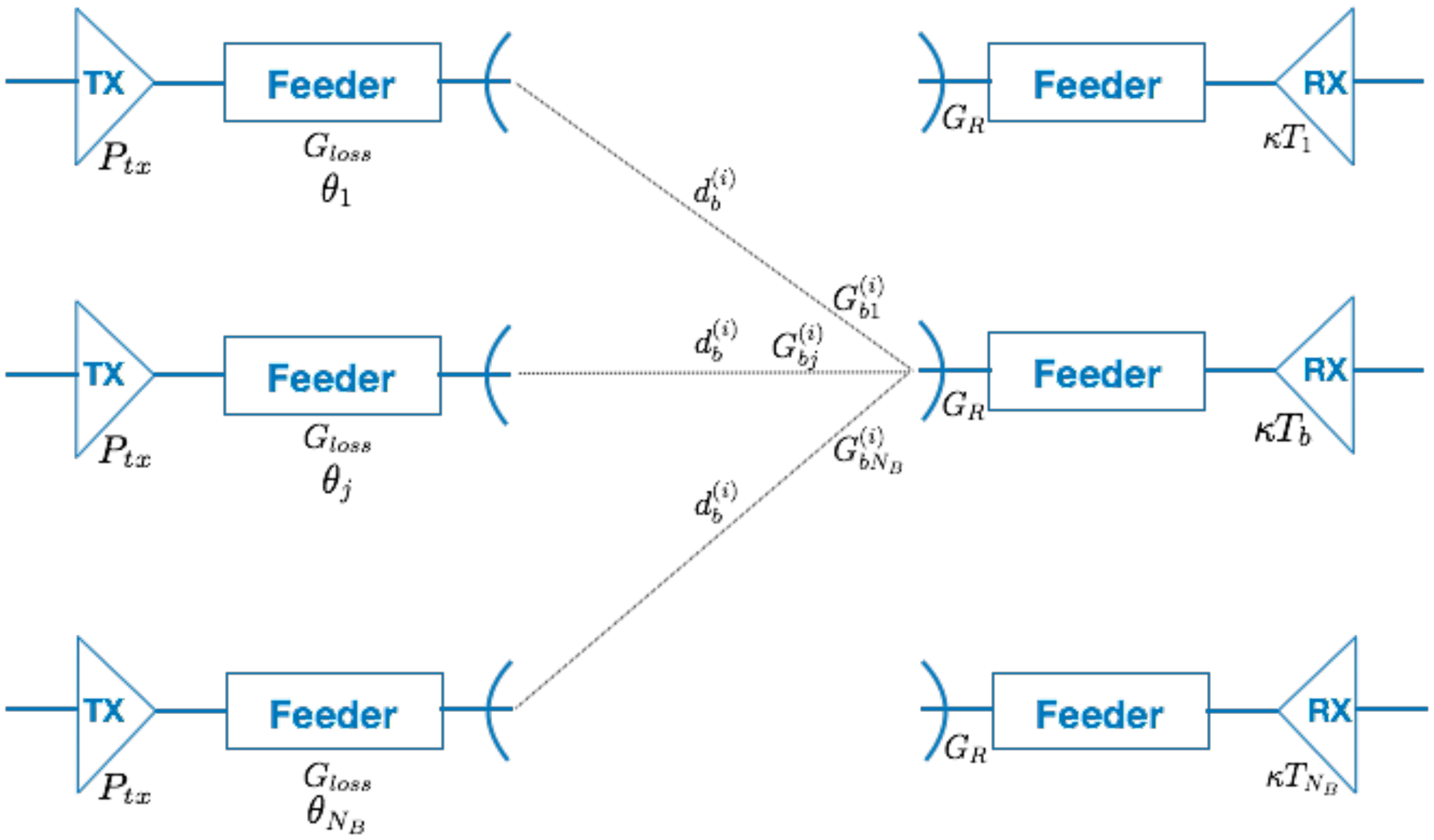}
\caption{Channel elements for the generic $i$-th user in the $b$-th beam.}
\label{fig:Channel}
\end{figure}

In Section~\ref{System model}, we introduce the multi-beam satellite system and multicast precoding model, as well as the problem statement. Section~\ref{sec:Random} outlines the traditional random scheduling algorithm. In Section~\ref{sec:GSA}, we introduce and describe the proposed Geographical Scheduling Algorithm. In Section~\ref{sec:Simulations}, numerical results are provided comparing the performance of the random and GSA algorithms. Finally, Section~\ref{sec:Conclusions} concludes this paper.

\section{System Model and Problem Statement}
\subsection{System model}
We consider a HTS GEO multi-beam satellite system operating with a full frequency reuse scheme, \cite{UniBo_Aerospace}. The satellite payload is assumed to be transparent and equipped with $N_B$ colocated transmitting antennas, generating $N_B$ on-ground beams. A single system Gateway (GW) manages the Channel State Information (CSI), which in the following is assumed to be ideal, obtained from the Return Link; in order to cope with the aggressive frequency reuse scheme, the CSI is exploited to compute the precoding weights by means of linear precoding techniques. In particular, Minimum Mean Square Error (MMSE) precoding is considered in the following. Time Division Multiple Access (TDMA) is implemented to serve the users in the $N_B$ on-ground beams in each time frame, as, for instance, in DVB-S2 and DVB-S2X, \cite{DVBS2,DVBS2X}.
\paragraph{User deployment and channel model} focusing on the generic $b$-th beam, we assume $N_U^{(b)} = \rho A_b$ uniformly distributed users with $A_b$ being the beam area and $\rho$ the user density, assumed to be the same across all beams. The $N_U^{(b)}$ users are further assumed to be in fixed locations. In this context, the channel coefficient between the generic $i$-th user of the $b$-th beam and the $j$-th transmitting antenna, with $i=1,\ldots,N_U^{(b)}$ and $i,j=1,\ldots,N_B$, is given by:
\begin{equation}
\label{eq:ChannelCoeff}
    h_{bj}^{(i)} = \frac{\sqrt{G_RG_{loss}G_{bj}^{(i)}}}{4\pi\frac{d_b^{(i)}}{\lambda}\sqrt{P_{Z,b}}}e^{-\jmath \frac{2\pi}{\lambda}d_b^{(i)}} e^{-\jmath \vartheta_b}
\end{equation}
where, as shown in Fig.~\ref{fig:Channel}: i) $G_R$ is the receiver antenna gain; ii) $G_{loss}$ models the overall antenna losses; iii) $G_{bj}^{(i)}$ is the multi-beam antenna gain between the $j$-th antenna feed and $i$-th user in the $b$-th receiving beam; iv) $d_b^{(i)}$ is the distance between the satellite and the considered $i$-th user in the $b$-th beam, which is the same for all colocated transmitting antennas; v) $\lambda$ the carrier wavelength; and vi) $\vartheta_b\sim\mathcal{U}\left[0,2\pi\right)$ is the random phase offset that depends on the transmitting antenna only. In addition, $P_{Z,b}=\kappa T_b B$ is the noise power at the $b$-th receiving antenna, in which $\kappa$ is the Boltzmann constant, $T_b$ the clear-sky noise temperature, and $B$ the user's bandwidth. In this context, the channel between the $i$-th user of the $b$-th beam and the $N_B$ transmitting antennas is given by $\mathbf{h}_b^{(i)} = \left(h_{b,1}^{(i)},\ldots,h_{b,N_B}^{(i)} \right)$ and the signal received on the Additive White Gaussian Channel (AWGN) can be written as:
\begin{equation}
\label{eq:RX_signal}
    y_b^{(i)} = \sqrt{p_b^{(i)}}\mathbf{h}_b^{(i)}\mathbf{x} + z_b^{(i)}, \ i=1,\ldots,N_U^{(b)}
\end{equation}
where: i) $\mathbf{x}$ is the $N_B\times 1$ vector of complex transmitted symbols; ii) $\sqrt{p_b^{(i)}}$ is the power allocated to the $i$-th user in the $b$-th beam\footnote{It shall be noted that this value is inherently different from the power emitted by each antenna, which is assigned by the precoder.}; and iii) $z_b^{(i)}$ is a complex circularly-symmetric independent and identically distributed (i.i.d.) Gaussian random variable with zero-mean and unit variance, since the noise term is included in the channel coefficients in (\ref{eq:ChannelCoeff}). In the following, we assume that $\sqrt{p_b^{(i)}}=\sqrt{P_{TX}}$, $\forall b,i$.

\paragraph{MMSE precoding} when traditional unicast precoding is implemented, one user per beam is selected based on the scheduling algorithm and a $N_B\times N_B$ MIMO channel matrix $\widetilde{\mathbf{H}} $ is built based on the CSI as $\widetilde{\mathbf{H}} = {\left( {\mathbf{h}_1^{(i)}}^T,\ldots,{\mathbf{h}_{N_B}^{(i)}}^T \right)}^T$, where, since we have one user per beam, we dropped the user index $i$ for the sake of clarity. This channel matrix is used to compute the MMSE precoding matrix as follows:
\begin{equation}
\label{eq:MMSE_PREC}
    \mathbf{W} = {\left(  \widetilde{\mathbf{H}}^H\widetilde{\mathbf{H}} + \mathrm{diag}\left(\mathbf{\bm{\alpha}}\right)\mathbf{I}_{N_B} \right)}^{-1}\widetilde{\mathbf{H}}^{H}
\end{equation}
where $\mathrm{diag}\left(\bm{\mathbf{\alpha}}\right)$ is the diagonal matrix of regularisation factors, with $\alpha_b = P_{Z,b}/P_{TX}$. The precoding matrix $\mathbf{W}$ is changed at each time frame, since the users to be served (\emph{i.e.}, the corresponding channel vectors) are different. The signal received in the precoded system can thus be written as:
\begin{equation}
    \mathbf{y} = \widetilde{\mathbf{H}}\widetilde{\mathbf{x}} + \mathbf{z} = \widetilde{\mathbf{H}}\mathbf{W}\mathbf{P}\mathbf{x} + \mathbf{z}
\end{equation}
where: i) $\widetilde{\mathbf{x}}=\mathbf{W}\mathbf{P}\mathbf{x}$ is the $N_B\times 1$ vector of precoded transmitted symbols; and ii) $\mathbf{P}$ is the diagonal matrix of allocated power levels $\mathbf{P}=\mathrm{diag}\left(\sqrt{p_1},\ldots,\sqrt{p_{N_B}}\right)=\sqrt{P_{TX}}\mathbf{I}_{N_B}$.\\
In SatCom systems, data sent to different users is multiplexed into a single codeword in each time frame and, in addition, their information bits are also interleaved together, making traditional symbol-level unicast precoding solutions unfeasible. To circumvent this issue, multicast precoding has been proposed in which the same precoding matrix is applied to all of the symbols in the same codeword. In this context, the precoding matrix $\mathbf{W}$ is constant over a time frame, and multiple users, denoted by $K$ in the following, are to be properly selected and multiplexed in the same codeword. Two main technical challenges arise: i) how to select the $K$ users to be multiplexed together in each beam, \emph{i.e.}, the clustering algorithm; and ii) how to process the users' channel vectors in each beam so as to have a single vector per beam. With respect to the latter, in \cite{Taricco}, average precoding has been proposed in which an equivalent channel matrix is built by means of a simple arithmetic. In particular, in the generic $b$-th beam, an equivalent channel vector is built as $\widetilde{\mathbf{h}}_b=\left(1/K\right)\sum_{i=1}^{K}\mathbf{h}_b^{(i)}$, which yields to the average estimated channel matrix $\widetilde{\mathbf{H}} = {\left( {\widetilde{\mathbf{h}}_1}^T,\ldots,{\widetilde{\mathbf{h}}_{N_B}}^T \right)}^T$. The more representative the equivalent channel vector $\widetilde{\mathbf{h}}_b$ is for the single users' channel vectors, the more adapted the precoding vector will be to the actual channel conditions.

\paragraph{Clustering algorithm} in \cite{UniBo_Aerospace}, the authors extensively discussed the problem of the user selection for multicast precoding by formulating it as a clustering problem. In particular, the generic $i$-th user in the $b$-th beam is represented by a feature vector $\mathbf{u}_i^{(b)}$, which depends on the similarity metric used in the clustering algorithm: i) its location in the Euclidean space when Euclidean distance is used as similarity; or ii) its vector of channel coefficients, $\mathbf{h}_b^{(i)}$, when users are grouped based on their distance in the $2N_B$-dimensional space of channel coefficients. Two algorithms were proposed, one with a fixed cluster size and another with variable cluster size, which showed improved performance with respect to SoA solutions. In the following, we assume that the MaxDist algorithm is implemented to cluster users together, which is mainly based on the following two steps, \cite{UniBo_Aerospace}: i) at each time frame, a reference user is selected as the farthest from the barycentre of the available users $\mathcal{Q}^{(b)}$\footnote{This represents the set of not yet clustered users.} in the considered similarity space, $\mathbf{g}^{(b)}$; and ii) a cluster is formed by taking the $K-1$ users in $\mathcal{Q}^{(b)}$ that are the closest to the reference user. This algorithm provides a $N_K^{(b)}$-partition of the $N_U^{(b)}$ users in the $b$-th beam, with $N_K^{(b)} = \left\lceil N_U^{(b)}/K\right\rceil$. This partition is the set of all of the users' cluster in the considered beam and it is represented by $\mathcal{C}^{(b)} = \left\{\mathcal{C}_1^{(b)},\ldots,\mathcal{C}_{N_K^{(b)}}^{(b)}\right\}$.
%

\begin{figure}[t]
\centering
\includegraphics[width=0.5\textwidth]{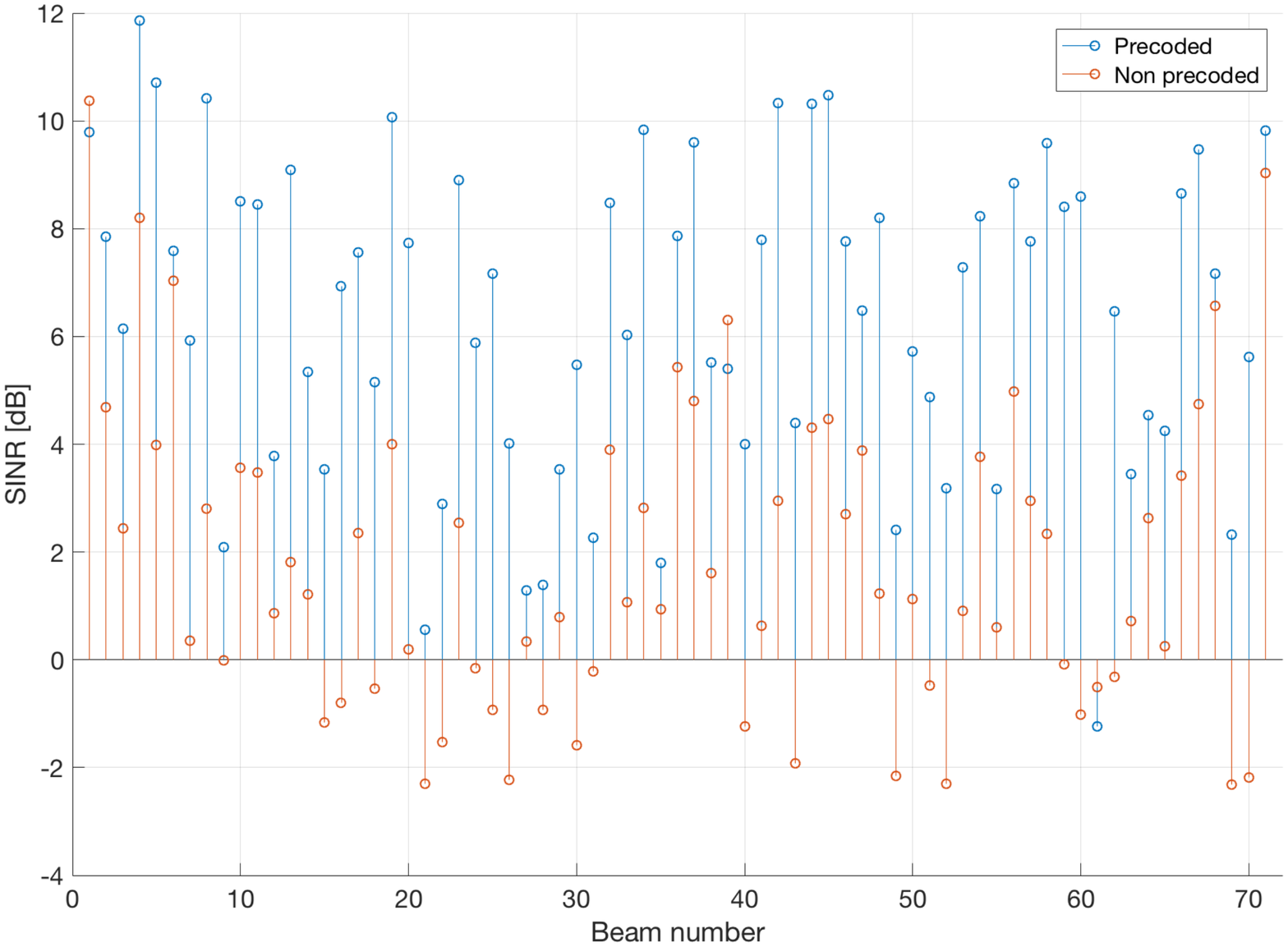}
\caption{SINR with and without precoding for $N_B=71$ receiving beams, MMSE precoding, and random scheduling.}
\label{fig:RandomStem}
\end{figure}
 
 \subsection{Problem statement}
In general, as shown in \cite{Taricco}-\cite{UniBo_Aerospace}, precoding provides a significant improvement in the overall system performance in terms of average spectral efficiency. However, even though the system average spectral efficiency is significantly improved, there are scenarios in which the Signal-to-Interference plus Noise Ratio (SINR) in the precoded system is below that of the non-precoded one for specific users. A situation like this is shown in Figure~\ref{fig:RandomStem} where in beams $1$, $36$, and $71$ (the details of the numerical simulations are provided in Section~\ref{sec:Simulations}) users are experiencing a loss in the system performance when precoding is implemented. This kind of situations occur when users in adjacent beams that are too closely located are scheduled in the same time frame; this happens due to the fact that the precoder cannot properly allocate the transmission power across the $N_B$ antennas so as to improve the intended signal and reduce the interfering ones, when users are too close. Typically, scheduling is implemented by means of a random algorithm , \emph{i.e.}, users are randomly selected across the system $N_B$ beams to be served in the same time frame, and scenarios in which the precoding performance is poor are quite frequent. In particular, approximately in $70\%$ of the transmitted time frames numerical results have shown at least one user for which the precoded SINR is below the non-precoded one. A similar behaviour can be observed also when multicast precoding is implemented. Based on these observations, in the following sections we introduce a Geographical Scheduling algorithm aimed at circumventing this issue.

\section{Random Scheduling}
\label{sec:Random}
In the considered SatCom multicast precoding system, one cluster per beam is selected from the partition $\mathcal{C}^{(b)} = \left\{\mathcal{C}_1^{(b)},\ldots,\mathcal{C}_{N_K^{(b)}}^{(b)}\right\}$ at each time frame to be served with TDMA. Let us denote by $n$, with $n\geq 1$, the generic time frame index. The set of clusters, one from each beam, served in the $n$-th time frame is denoted by $\mathcal{S}[n]=\left\{s_1[n], \ldots, s_{N_B}[n]\right\}$, in which the generic $b$-th element $s_b[n]$ represents the index of the cluster selected for the $b$-th beam. At the first time frame $n=1$, in each beam, all of the clusters are available for selection, due to the fact that none of them has been previously served. This means that $s_b[1]\in\left[1,N_K^{(b)}\right], \forall b$. In the generic time frame $n>1$, the system scheduler shall select a new cluster to be served from each beam, with the constraint that those already served in previous time frames are not eligible. Thus, in the generic time frame $n\geq 1$, the scheduler shall select the cluster to be served in the generic $b$-th beam from the following set:
\begin{equation}
\label{eq:AvailableIndex}
    \mathcal{A}_b[n]=\left\{1,\ldots,K_b\right\}\setminus\bigcup_{m=1}^{n-1}s_b[m]
\end{equation}
It is straightforward to note that $\mathcal{A}_b[1]=\left\{1,\ldots,K_b\right\}$, $\forall b$. 

When focusing on precoded SatCom systems, the scheduling algorithm is typically assumed to be random. In particular, in the generic $n$-th time frame the cluster to be served, the index of which was denoted as $s_b[n]$, in the $b$-th beam is randomly chosen from those that are still available, \emph{i.e.}, $s_b[n]\sim \mathcal{U}\left(\mathcal{A}_b[n]\right)$. This scheduling algorithm is referred to as \emph{Random Scheduling} in the following and it is outlined in Algorithm~\ref{alg:Random} for the generic $b$-th beam. It shall be noticed that each beam has been partitioned in $N_K^{(b)}$ clusters, \emph{i.e.}, the number of clusters is not constant across the system beams. To cope with this aspect, if the randomly chosen cluster (Step 2) is the last one available in the beam, the scheduler shall re-initialise the pool of available clusters $\mathcal{A}_b$ to $\mathcal{A}_b[1]$ (Step 6). This condition is required so as to guarantee that, while still serving clusters in the more populated beams, we have users also in the beam with less clusters. If there are still unserved clusters in the considered beam, the pool of available clusters for the next iteration is updated based on eq.~(\ref{eq:AvailableIndex}) (Step 4). The overall scheduling sequence for the random algorithm is finally given by $\mathcal{S}_{RAND}=\bigcup_{n=1}^{N_{frame}}\mathcal{S}[n]$.
 \begin{algorithm}
\caption{Random scheduling algorithm}
\label{alg:Random}
\begin{algorithmic}[1]
\renewcommand{\algorithmicrequire}{\textbf{Input:}}
\renewcommand{\algorithmicensure}{\textbf{Output:}}
    \REQUIRE $N_K^{(b)}$-partition $\mathcal{C}^{(b)}=\left\{\mathcal{C}_1^{(b)},\ldots,\mathcal{C}_{N_K^{(b)}}^{(b)}\right\}$
    \ENSURE  $\mathcal{S}[n]=\left\{s_1[n], \ldots, s_{N_B}[n]\right\}$, $1\leq n\leq N_{frame}$ \\
    \STATE Set $\mathcal{A}_b[1]=\left\{1,\ldots,N_K^{(b)}\right\}$, $N_{frame}\geq \max_b\left\{N_K^{(b)}\right\}$
    \FOR {$n=1$ to $N_{frame}$}
        \FOR {$b=1$ to $N_B$ do}
        \STATE Cluster random selection: $s_b[n]\sim \mathcal{U}\left(\mathcal{A}_b[n]\right)$
  \IF {$\left(n< N_K^{(b)}\right)$}
  \STATE Update available clusters: $\mathcal{A}_b[n+1]=\mathcal{A}_b[n]\setminus s_b[n]$
  \ELSE
  \STATE Re-initialise available clusters: $\mathcal{A}_b[n+1]=\mathcal{A}_b[1]$
  \ENDIF
  \ENDFOR
 \RETURN $\mathcal{S}[n]$
 \ENDFOR
\end{algorithmic}
\end{algorithm}

\section{Geographical Scheduling}
\label{sec:GSA}
In this section, we propose a Geographical Scheduling Algorithm (GSA) aimed at circumventing the issues identified in Section~\ref{sec:Random}. In particular, the algorithm design principle is that of scheduling clusters in adjacent beams in the same time frame only if they are not too close. A possible solution would be that of selecting these two clusters so as to have the maximum distance between them. However, this would lead to the same critical scenario as the one we are trying to avoid, as these two cluster might then be too close to the clusters in the other adjacent beams. To circumvent this issue, while scheduling together clusters that are not too close to each other, we propose to define a certain number of zones within each beam, which will be referred to as \emph{scheduling sectors}, and only schedule together clusters that belong to the same scheduling zone in their respective beams.

\begin{algorithm}
\caption{Geographical Scheduling Algorithm}
\label{alg:Geo1}
\begin{algorithmic}[1]
\renewcommand{\algorithmicrequire}{\textbf{Input:}}
\renewcommand{\algorithmicensure}{\textbf{Output:}}
    \REQUIRE $\mathcal{Z}_{q}^{(b)}, N_K^{(b)}$-partition $\mathcal{C}^{(b)}$
    \ENSURE  $\mathcal{S}^{(q)}[n_q]=\left\{s_1^{(q)}[n_q], \ldots, s_{N_B}^{(q)}[n_q]\right\}$, $1\leq n_q\leq N_{frame}^{(q)}$ \\
    \STATE Set $\mathcal{A}_{b}^{(q)}[1]=\left\{i\left| \mathbf{v}_i^{(b)}\in  \mathcal{Z}_{BC}^{(q)}\right.\right\}$
    \FOR {$n_q=1$ to $N_{frame}^{(q)}$}
    \FOR {$b=1$ to $N_B$ do}
        \STATE Cluster selection in sector $q$: $s_b^{(q)}[n_q]\sim\mathcal{U}\left( \mathcal{A}_{b}^{(q)}[n_q]\right)$
    \IF {$\left(\left|  \mathcal{A}_{b}^{(q)}[n_q] \right|>1\right)$}
      \STATE Update available clusters in sector $q$: $\mathcal{A}_{b}^{(q)}[n+1]=\mathcal{A}_{b}^{(q)}[n]\setminus s_b^{(q)}[n]$
      \ELSE
      \STATE Re-initialise available clusters: $\mathcal{A}_{b}^{(q)}[n+1]=\mathcal{A}_{b}^{(q)}[1]$
    \ENDIF
    \ENDFOR
 \RETURN $\mathcal{S}^{(q)}[n_q]$
 \ENDFOR
\end{algorithmic}
\end{algorithm} 

\subsection{Sectorisation}
Focusing on a generic beam, the aim is to define $N_{S}$ scheduling sectors to be exploited by the GSA. The first zone that we identify is the Beam Center sector, $\mathcal{Z}_{BC}$, which can be defined as the locus of points the distance of which from the beam center is not exceeding a pre-defined beam center radius. Since the beams are not circular and, actually, show a different geometry one from another, we introduce a normalised polar coordinate system centered in the beam center. In particular, denoting by $\mathbf{v}_i^{(b)}$ the location of the generic $i$-th user, we identify its location by means of its: i) angular coordinate, $\varphi\left(\mathbf{v}_i^{(b)}\right)$; and ii) its normalised radius $\widetilde{r}\left(\mathbf{v}_i^{(b)}\right)$, \emph{i.e.}, its distance from the beam center normalised to the beam edge distance in the direction $\varphi\left(\mathbf{v}_i^{(b)}\right)$:
\begin{equation}
\label{eq:NormRadius}
    \widetilde{r}\left(\mathbf{v}_i^{(b)}\right) = \widetilde{r}_i^{(b)}=\frac{\left|\mathbf{v}_i^{(b)}\right|}{R\left(\varphi\left(\mathbf{v}_i^{(b)}\right)\right)}
\end{equation}
By construction, it can be easily verified that $0\leq \widetilde{r}_i^{(b)} \leq 1$: when the selected point $\mathbf{v}_i^{(b)}$ coincides with the beam center, the distance $\left|\mathbf{v}_i^{(b)}\right|=0$ in (\ref{eq:NormRadius}) and $\widetilde{r}_i^{(b)}=0$, while, if the point is located at beam edge, then we have $\left|\mathbf{v}_i^{(b)}\right|=R\left(\varphi\left(\mathbf{v}_i^{(b)}\right)\right)$ and $\widetilde{r}_i^{(b)}=1$. Based on the normalised radius, we can define the beam center sector $\mathcal{Z}_{BC}$ as the locus of points with distance from the beam center that is lower than or equal to a predefined normalised radius, $r_{BC}$:
\begin{equation}
\label{eq:BeamCenterSector}
    \mathcal{Z}_{BC}^{(b)} = \left\{ \mathbf{v}_i^{(b)}\in\mathcal{B}\left| \widetilde{r}_i^{(b)} \leq r_{BC}    \right. \right\}
\end{equation}
where $\mathcal{B}_b$ is the locus of points belonging to the $b$-th beam.

\begin{figure}[t]
\centering
\includegraphics[width=0.5\textwidth]{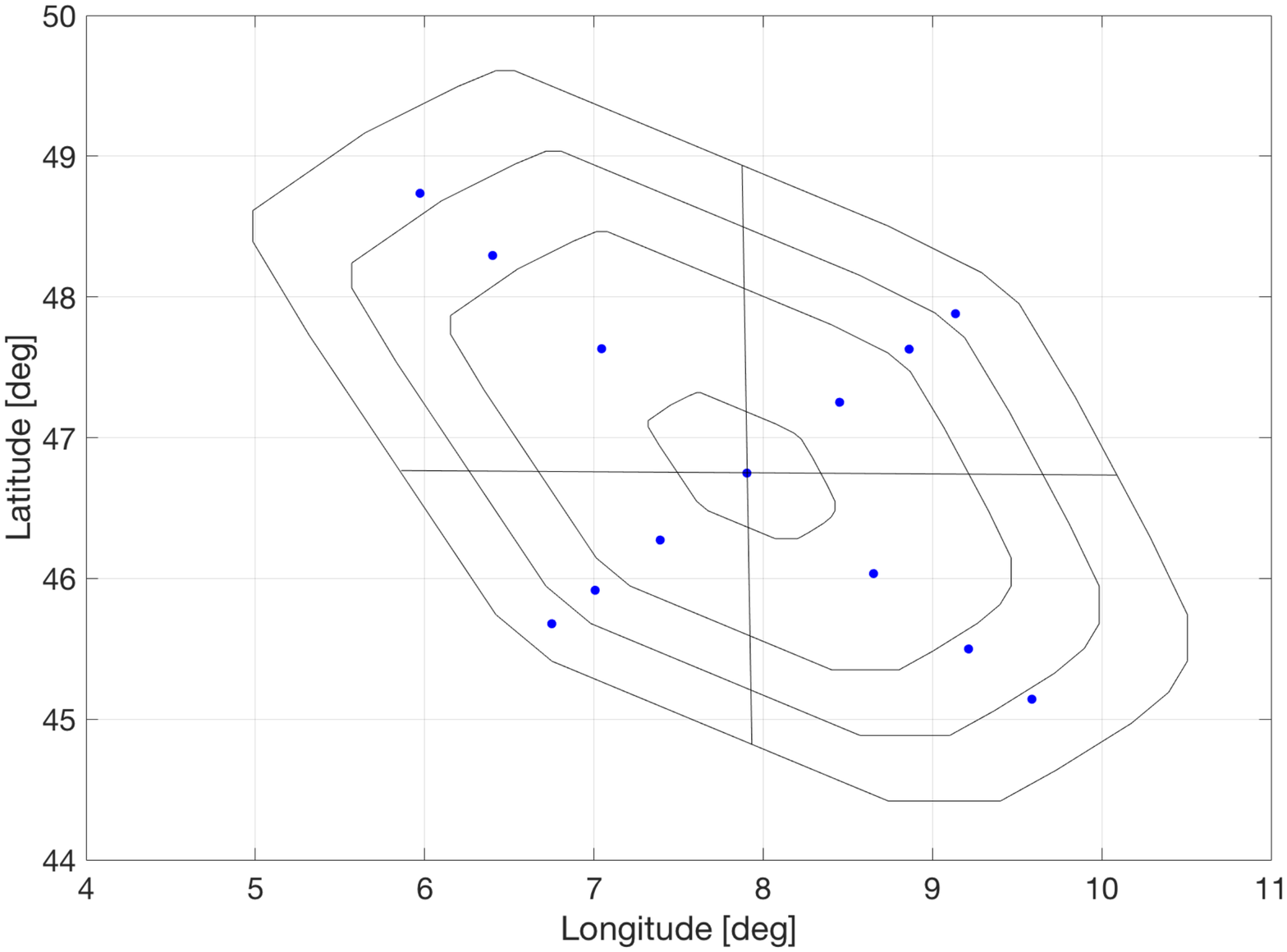}
\caption{Sectorisation example for beam $42$ with $N_S=13$. Setup: $r_{BC}=0.2$, $r_1=0.6$, $r_2=0.8$, $r_3=1$; $\varphi_0=0$, $\varphi_1=\pi/2$, $\varphi_2=\pi$, $\varphi_3=2\pi$.}
\label{fig:Sectors}
\end{figure}

In order to define the other scheduling sectors, we identify: i) $N_{R}$ normalised radius values, in addition to $r_{BC}$, to discern among sectors based on their distance from the beam center; and ii) $N_{\varphi}$ angles to discern among sectors based on the direction we are observing from the beam center. Thus, the number of sectors identified with the proposed approach is given by $N_S=N_R\cdot N_{\varphi}+1$, in which we must sum $1$ to include the previously defined beam center sector $\mathcal{Z}_{BC}^{(b)}$. The $q$-th scheduling sector $\mathcal{Z}_q^{(b)}$ in the $b$-th beam is defined as the locus of points having a normalised radius in the range $\left(r_{q-1},r_q\right]$ and falling in an angular range $\left(\varphi_{q-1},\varphi_{q}\right]$:
\begin{equation}
\label{eq:SchedulingSectors}
    \mathcal{Z}_q^{(b)}=\left\{ \mathbf{v}_i^{(b)}\in\mathcal{B}_b \left| r_{q-1}<\widetilde{r}_i^{(b)}\leq r_q, \varphi_{q-1}<\varphi_{\mathbf{v}_i^{(b)}}\leq \varphi_{q}  \right. \right\}
\end{equation}
with $r_0=r_{BC}$, $\mathcal{Z}_{BC}^{(b)}\bigcup\left\{\bigcup_{q=1}^{N_Z}\mathcal{Z}_q^{(b)}\right\}=\mathcal{B}_b$, $\varphi_0=0$, and $\varphi_{N_S-1}=2\pi$. It shall be noted that the sectors boundaries, and in particular their radius identified by the range $\left(r_{q-1},r_q\right]$, are not depending on the considered beam, since they are adapted to its geometry by construction being normalised values. Finally, it is worth noting that the above identification of the sector in which a user is falling can be easily extended to the considered multicast scenario. In particular, each cluster is assumed to be represented by its centroid, \emph{i.e.}, barycenter, which is then used to identify its sector based on (\ref{eq:BeamCenterSector})-(\ref{eq:SchedulingSectors}). Figure~\ref{fig:Sectors} shows an example of beam sectors for beam $42$, in which the sector barycentres are highlighted.

\subsection{Geographical Scheduling Algorithm (GSA)}
\label{sec:GSA}
The proposed GSA scheduling is outlined in Algorithm~\ref{alg:Geo1}. Once the beam sectors have been defined as described above, the scheduling algorithm only schedules together clusters across the $N_B$ system beams that are located in the same sector of their beam. Without any loss of generality, let us assume that the geographical scheduler serves the users in increasing orders of scheduling sectors, \emph{i.e.}, first we serve the clusters belonging to $\mathcal{Z}_{BC}^{(b)}$ and then $\mathcal{Z}_{q}^{(b)}$ for $q=1,\ldots,N_S-1$. The GSA first identifies the cluster indexes that are related to clusters in the considered $q$-th sector (Step 1), denoted by $\mathcal{A}_b^{(q)}$, and uses it as the initial pool of available clusters to be scheduled. At each time frame, the scheduler randomly select a cluster to be served from the pool of those that have not been served yet, \emph{i.e.}, $s_b^{(q)}[n_q]\sim\mathcal{U}\left( \mathcal{A}_{b}^{(q)}[n_q]\right)$ (Step 3). The time frame index $n_q$ ranges from $1$ to the maximum number of frames to be transmitted for that sector, which is given by the maximum number of clusters in the $q$-th sector across all beams: $N_{frame}^{(q)}=\max_b\left\{\left| \mathcal{Z}_{q}^{(b)} \right|\right\}$. This condition is necessary so as to guarantee that all of the clusters belonging to the $q$-th sector are served at least once in all beams. Moreover, this upper bound on $n_q$ implies that we transmit $N_{frame}=\sum_{q=1}^{N_S-1}N_{frame}^{(q)} + N_{frame}^{(BC)}$ frames when serving clusters in all sectors and beams. In addition, please note that, for the beam center sector $\mathcal{Z}_{BC}^{(b)}$, we can simply substitute the index $q$ with $BC$ in Algorithm~\ref{alg:Geo1}.\\
Similarly to the Random algorithm, the set of clusters available for scheduling is continuously updated (Step 3), which leads to the following definition of the $q$-th sector\footnote{$q=BC$ for the beam center.} set of available users in the generic time frame $n_q$:
\begin{equation}
    \mathcal{A}_b^{(q)} = \left\{s_1^{(q)}[n_q], \ldots, s_{N_B}^{(q)}[n_q]\right\}\setminus \bigcup_{m=1}^{n_q}s_b^{q}[m]
\end{equation}
By repeating Algorithm~\ref{alg:Geo1} for all sectors, we obtain the overall scheduling sequence:
\begin{equation}
\mathcal{S}_{GSA} = \bigcup_{n_{BC=1}}^{N_{frame}^{(BC)}} \mathcal{S}^{(BC)}[n_{BC}] \bigcup \left\{\bigcup_{q=1}^{N_S-1}\bigcup_{n_q=1}^{N_{frame}^{(q)}} \mathcal{S}^{(q)}[n_{q}]\right\}
\end{equation}

%

\section{Numerical Results}
\label{sec:Simulations}
In this section, we assess compare the performance of the proposed GSA with the random scheduling algorithm in terms of average spectral efficiency. The main simulation parameters are listed in Table~\ref{tab:SimulationParam} and the considered multi-beam satellite system covers the whole of Europe through $N_B=71$ beams operating with a full frequency reuse scheme. As for the user densities across the beams, the largest value is $10^{-2}$ users/km$^2$, which is a low density. While this choice is motivated by the memory and time computational complexity, numerical results show that it is sufficient to understand the trend for the overall system performance. Based on this density, during each Monte Carlo iteration, $N_U^{(b)}=\left[\rho A_b\right]$ users are randomly deployed in fixed locations in each beam with a uniform distribution. Finally, we assume a uniform traffic request for the users, \emph{i.e.}, all users are requesting the same amount of traffic and no priorities are present or requested, and the multicast precoding clusterisation based on the channel coefficient space. 
 
For the generic $c$-th cluster in the $b$-th beam and $n$-th time frame, we obtain a rate $\eta_{c,n}^{(b)}\left(K,\rho\right)$, which is a function of the minimum SINR among the cluster members, since all of them shall be in condition to decode the received data, \emph{i.e.}:
\begin{equation}
\label{eq:NumericalRate}
    \eta_{c,n}^{(b)}(K,\rho) = f\left(\widetilde{\gamma}_c^{(b)}\right),\ \widetilde{\gamma}_c^{(b)}=\min_{i\in\mathcal{C}_c^{(b)}}\left\{\gamma_{c,i}^{(b)}\right\}
\end{equation}
The function $f(\cdot)$ models the considered Modulation and Coding (ModCod) scheme, which in the following is assumed to be the one provided by the DVB-S2X standard with $64800$ bits FEC codewords, \cite{DVBS2X}. Thus, the average spectral efficiency is computed by averaging over all time frames, clusters, and beams:
\begin{equation}
\label{eq:AverageRate}
    \overline{\eta}(K,\rho) = \mathbb{E}_{n,c,b}\left\{\eta_{c,n}^{(b)}(K,\rho)\right\}
\end{equation}

\begin{table}[t]
\renewcommand{\arraystretch}{1.3}
\caption{Numerical simulation parameters}
\label{tab:SimulationParam}
\centering
\begin{tabular}{|c|c|}
\hline
\bfseries Parameter & \bfseries Value\\
\hline\hline
Carrier frequency & $19.5$ GHz\\
\hline
Receiving antenna diameter & $0.6$ m\\
\hline
Receiving antenna efficiency & $0.6$\\
\hline
Antenna losses & $2.55$ dB\\
\hline
GEO satellite longitude & $30^{\circ}$\\
\hline
Satellite transmitted power & $P_{sat}=90$W\\
\hline
$\rho$ & $2.5\cdot 10^{-4}, \ldots, 10^{-2}$ users/km$^2$\\
\hline
$K$ & $1$ (unicast), $2,4,\ldots,12$\\
\hline
Target Bit Error Probability & $10^{-5}$\\
\hline
Sector radii & $r_{BC}=0.2$, $r_1=0.6$, $r_2=0.8$, $r_3=1$\\
\hline
Sector angles & $\varphi_0=0$, $\varphi_1=\pi/2$, $\varphi_2=\pi$, $\varphi_3=2\pi$\\
\hline
\end{tabular}
\end{table}

\begin{figure}[t]
\centering
\includegraphics[width=0.5\textwidth]{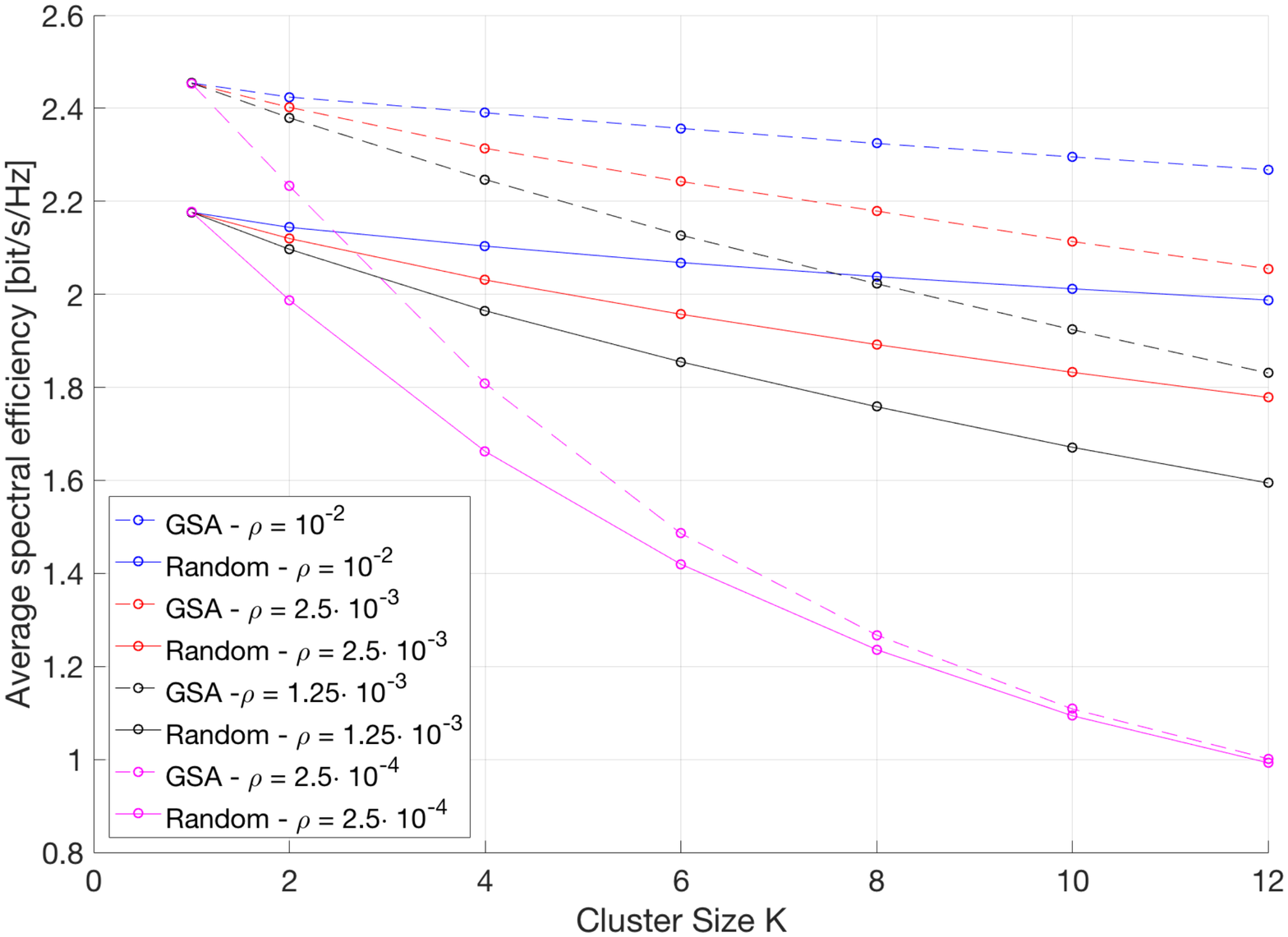}
\caption{Average spectral efficiency with the random and GSA algorithms as a function of the cluster size $K$ for varying user densities $\rho$.}
\label{fig:RATE}
\end{figure}

Figure~\ref{fig:RATE} shows the average spectral efficiency with users' clustering based on the channel coefficient similarity with the random and GSA scheduling algorithms. The overall trend related to the increase in the cluster size is the same for both scheduling algorithms and reflects the analysis provided in \cite{UniBo_Aerospace}. In particular:
\begin{itemize}
    \item the average spectral efficiency decreases for increasing values of the cluster size. This is due to the fact that the precoding matrix in (\ref{eq:MMSE_PREC}) is built by averaging the cluster users' channel coefficients and, thus, the precoding matrix will be less representative. This effect is more evident in low density scenarios, since users are more spread in the considered similarity space.
    \item Typically, the channel coefficient similarity provides better performance. This is due to the fact that users that are close in the Euclidean space do not necessarily have similar channel coefficients.
\end{itemize}

\begin{figure}[t]
\centering
\includegraphics[width=0.5\textwidth]{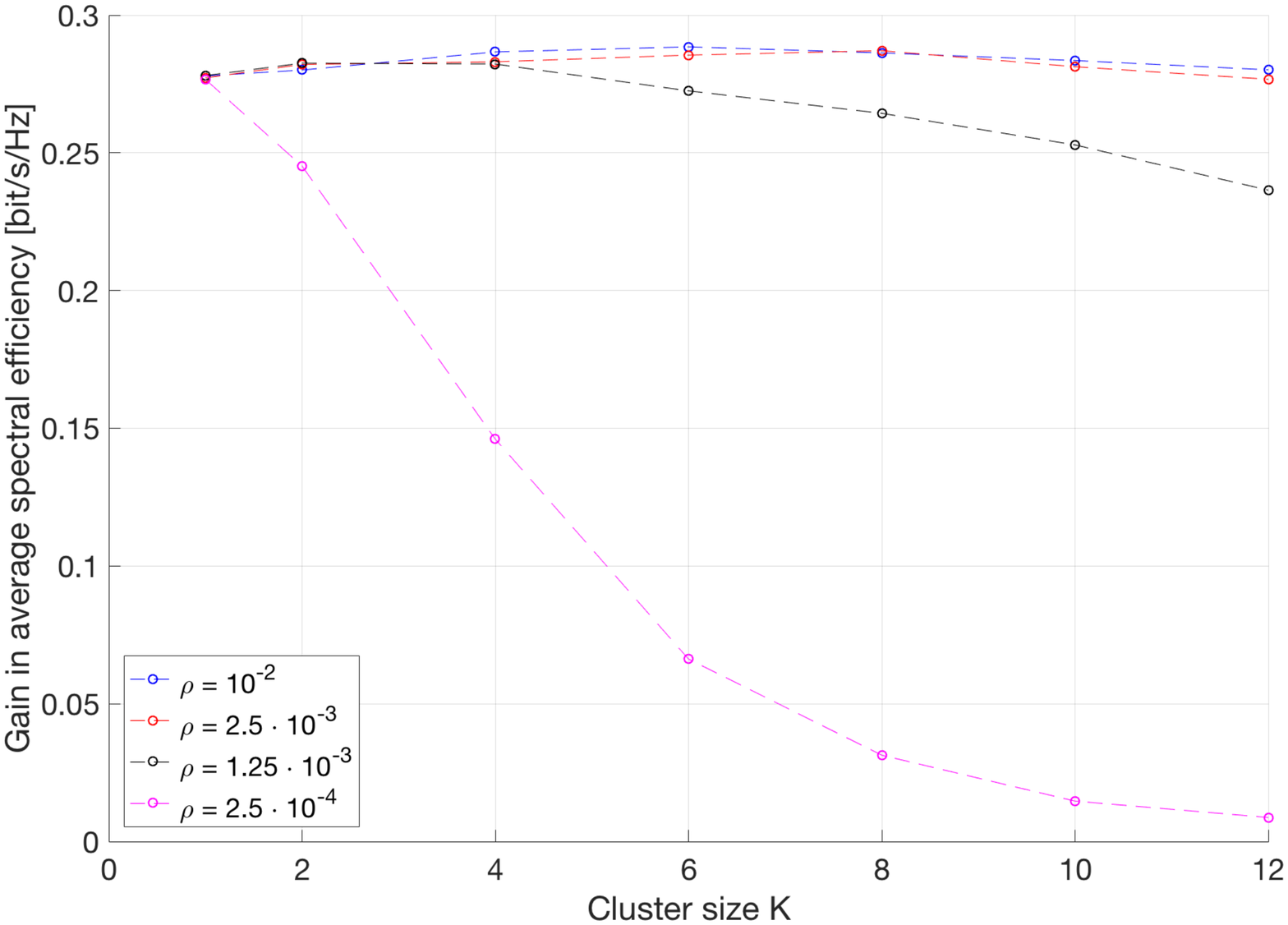}
\caption{Gain in average spectral efficiency with the GSA w.r.t. the random scheduling algorithm as a function of the cluster size $K$ for varying user densities $\rho$.}
\label{fig:RATE_GAIN}
\end{figure}

With respect to the proposed GSA solution, it can be noticed that the performance gain with respect to the random scheduling is significant. This aspect is highlighted in Figure~\ref{fig:RATE_GAIN}, where the gain in average spectral efficiency is shown, \emph{i.e.}, the average spectral efficiency with the GSA algorithm with respect to the one obtained with the random scheduling. It can be noticed that with the lowest user density, the gain rapidly decreases to almost zero for increasing user densities. This is due to the fact that we are clustering together more users in a scenario in which they are very sparse in the beam area. Consequently, while the barycentre of each cluster will denote its sector, the cluster users will be far away from each other, which leads back to the critical scenarios in the random clustering. On the other hand, for larger user densities, the average gain is quite stable and included between $0.25$ and $0.3$ bit/s/Hz. It can still be noticed that for an increasing cluster size the gain tends to decrease, with the same motivation as for the lowest density. However, in this case the gain would go to zero only for extremely large cluster sizes.

\begin{figure}[t]
\centering
\includegraphics[width=0.5\textwidth]{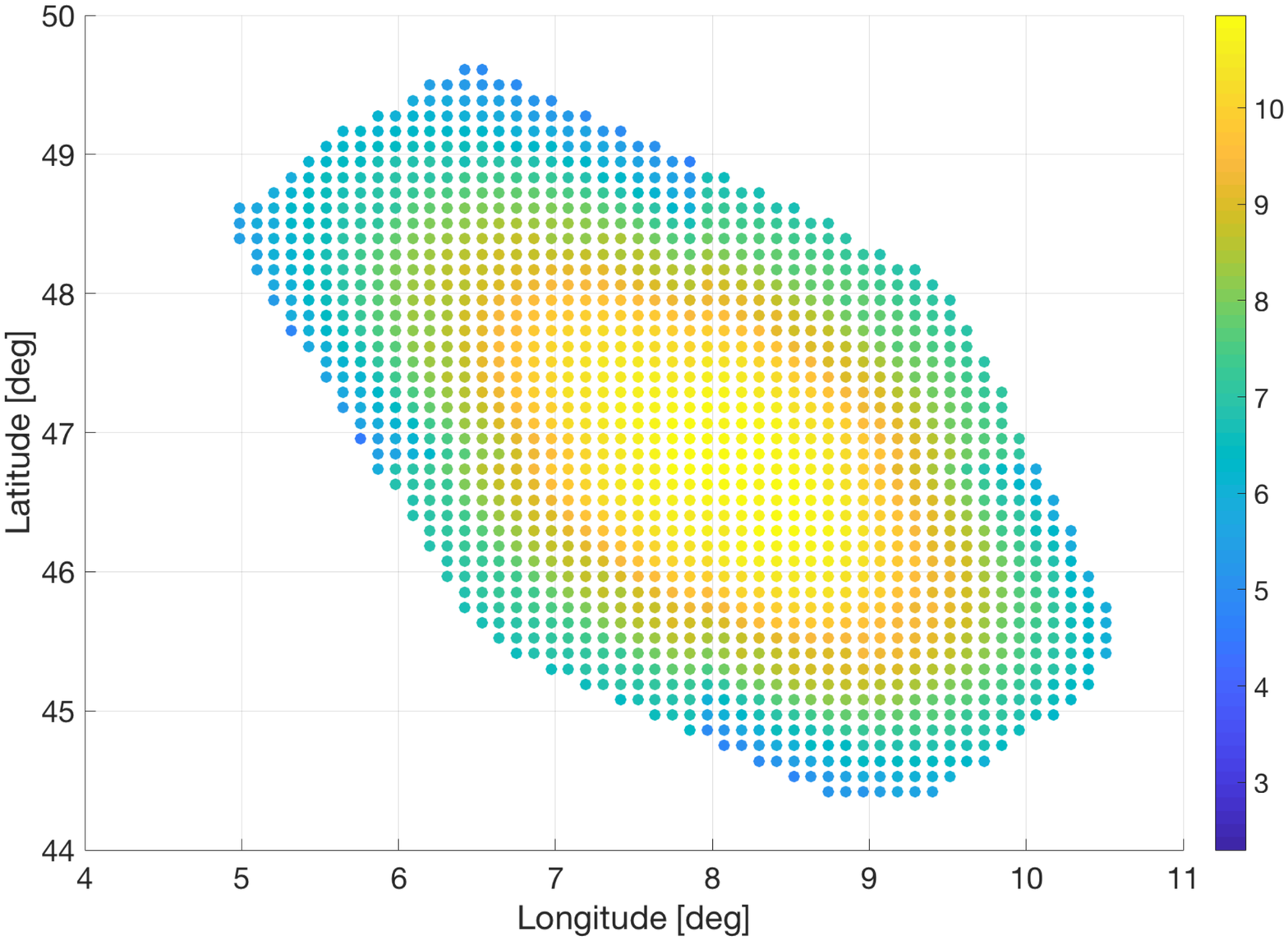}
\caption{Average SINR per user with the GSA algorithm. Setup: beam $42$, $\rho=2.5\cdot 10^{-3}$ users/km$^2$.}
\label{fig:MAP_GSA}
\end{figure}

\begin{figure}[t]
\centering
\includegraphics[width=0.5\textwidth]{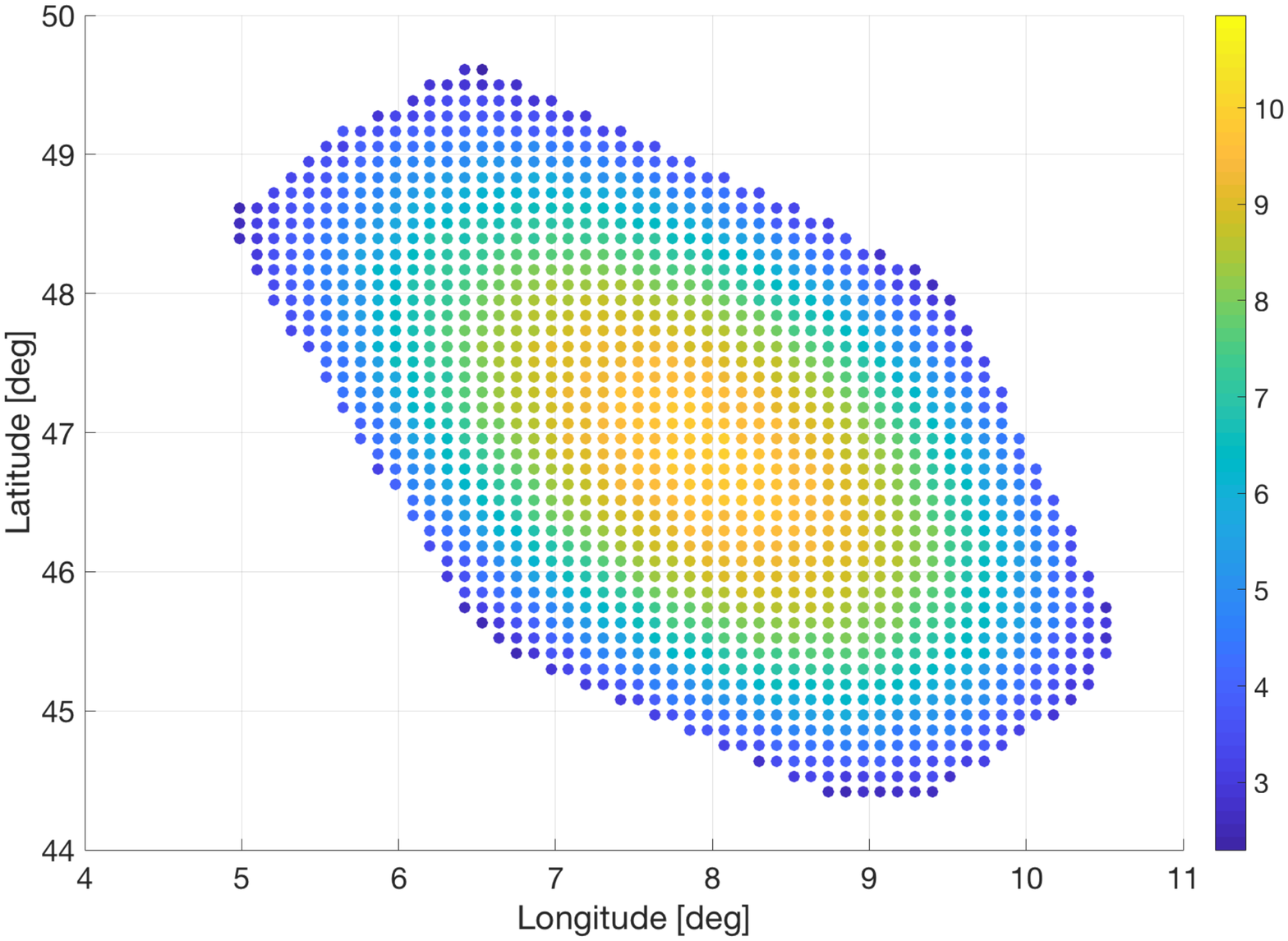}
\caption{Average SINR per user with the random scheduling algorithm. Setup: beam $42$, $\rho=2.5\cdot 10^{-3}$ users/km$^2$.}
\label{fig:MAP_RANDOM}
\end{figure}

This gain is not only obtained thanks to avoiding critical scenarios, \emph{i.e.}, scenarios in which users from adjacent beams and close to each other are scheduled in the same time frame. Figures \ref{fig:MAP_GSA}-\ref{fig:MAP_RANDOM} shows the average SINR for beam $42$ with the GSA and random algorithms, respectively. It can be noticed that even users that were distant from the beam edge are experiencing a significantly improved SINR with the proposed GSA. Moreover, the average SINR is more uniform with the GSA, which shows that this solution not only provides improved an average spectral efficiency at system level, but it also improves the overall fairness for the users.

\section{Conclusions}
\label{sec:Conclusions}
In this paper, we designed a Geographical Scheduling Algorithm (GSA) for multicast precoding in multi-beam HAT satellite systems. This algorithm is based on the concept that, by scheduling together users belonging to similar locations in their respective beams (denoted as beam sectors), the precoding performance are significantly improved. In fact, when users in adjacent beams and closely located are scheduled in the same time frame, the precoding performance is poor since the precoder cannot find a proper power allocation to transmit their information. With the proposed GSA, these situations are avoided and, in addition, also users that were experiencing already good average spectral efficiency levels can see a significant performance benefit. In addition, it has been shows that the proposed GSA tends to uniform the SINR levels throughout the beam, which leads to an improved fairness in the system performance.

\section*{Acknowledgment}
This work has been supported by European Space Agency (ESA) funded activity SatNEx IV CoO2-PART 1 WI4 ``Forward Packet Scheduling Strategies for Emerging Satellite Broadband Networks.'' Opinions, interpretations, recommendations and conclusions presented in this paper are those of the authors and are not necessarily endorsed by the European Space Agency.

\end{document}